\documentclass{PoS}

\title{Hunting for heavy Dark Matter in the Galactic Center with ground-based Cherenkov telescopes}

\ShortTitle{Hunting for heavy DM in the GC with Cherenkov telescopes}

\author{\speaker{Lucia Rinchiuso}\\
        IRFU, CEA, Universit\'e Paris-Saclay, F-91191 Gif-sur-Yvette, France\\
        E-mail: \email{lucia.rinchiuso@cea.fr}}

\author{Nicholas L. Rodd\\
        Berkeley  Center  for  Theoretical  Physics,  University  of  California,  Berkeley,  CA  94720,  USA\\
        Theoretical  Physics  Group,  Lawrence  Berkeley  National  Laboratory,  Berkeley,  CA  94720,  USA\\
        E-mail: \email{nrodd@berkeley.edu}}

\author{Emmanuel Moulin\\
        IRFU, CEA, Universit\'e Paris-Saclay, F-91191 Gif-sur-Yvette, France\\
        E-mail: \email{emmanuel.moulin@cea.fr}}

\author{Tracy R. Slatyer\\
        Center  for  Theoretical  Physics,  Massachusetts  Institute  of  Technology,  Cambridge,  MA  02139,  USA\\
        School of Natural Sciences Institute for Advanced Study, Princeton, NJ 08540, USA\\
        E-mail: \email{tslatyer@mit.edu}}

\abstract{A TeV scale electroweak particle is a well motivated candidate for the dark matter (DM) of our Universe. Yet such a particle may only be detectable using indirect detection instruments sensitive to TeV-scale gamma rays that can result from dark matter annihilations. We present a mock analysis of the sensitivity for the present ground-based Cherenkov telescope array H.E.S.S. (High Energy Spectroscopic System) to detect TeV scale DM in the Galactic Center region. The work combines next-to-leading-logarithmic order calculations for the annihilation photon spectrum, as well as a comprehensive treatment of detector effects and expected backgrounds. Forecast limits on the sensitivity of H.E.S.S. have been derived across the important TeV mass range, assuming different DM density profiles and focusing on the canonical WIMP dark matter candidate Wino. These limits test our present and future ability to probe the predicted thermal cross section for some of the most promising DM candidates that could be discovered in the coming decade.}

\FullConference{36th International Cosmic Ray Conference -ICRC2019-\\
		July 24th - August 1st, 2019\\
		Madison, WI, U.S.A.}

\begin{document}

\section{Introduction}
About 84\% of the matter in the Universe is made out of dark matter (DM) whose nature is still unknown. A good elementary particle candidate for DM is expected to be electrically neutral, colorless, non-baryonic, cold in the standard cosmological model and stable over the scale of the Universe. 
Dark matter can be detected indirectly through the detection of the final products of its self-annihilation among which very-high-energy (VHE) gamma rays can be found. The VHE gamma rays travel straight from the source and initiate electromagnetic showers in the Earth atmosphere. The charged particles in the shower are relativistic enough to produce Cherenkov light that can be detected by ground-based Imaging Atmospheric Cherenkov telescopes (IACTs) that fall in the light pool. 

The High Energy Spectroscopic System (H.E.S.S.) is an array of five IACTs set in Namibia. It consists of four telescopes with diameter 12~m (H.E.S.S. I) and a fifth bigger telescope of diameter 28~m (H.E.S.S. II) added in 2012 in the middle of the array.  H.E.S.S. can detect VHE gamma rays in the energy range of about 100~GeV-100~TeV  with an energy resolution of 10\% and an angular resolution better than $0.1^\circ$. Being located in the Souther hemisphere H.E.S.S. is in an optimal position to observe the Galactic Center (GC). With H.E.S.S.~I about 250 hours of data have been collected towards the GC. The inner Galactic halo is also one of the densest DM regions in the Universe and a very promising target for indirect DM search due to its short distance from us. H.E.S.S. has been very active in the DM searches towards the GC~\cite{Abdallah:2018qtu, Abdallah:2016ygi} providing the most stringent constraints so far on the DM annihilation cross section in the TeV DM mass range.

\section{The Wino dark matter candidate}

The Wino represents a compelling DM candidate to hunt for with IACTs for two reasons.
Firstly, there are a number theoretical arguments that suggest the Wino could well be the particle that DM is made of.
If this is the case, IACTs could be on the verge of discovery.
Secondly, the indirect detection signature of the Wino involves a number of effects that significantly deviate from simple estimates.
Yet these effects are expected to be quite generic for heavy DM candidates, and therefore the Wino represents an opportunity to determine the impact of such effects on the experimental sensitivity.

The Wino is an exceptionally simple DM candidate: a triplet of Majorana fermions charged only under SU(2)$_L$, where it transforms in the adjoint.
After electroweak symmetry breaking, the triplet splits into a singly charged Dirac fermion, and a slightly lighter neutral particle which represents the DM.
This simplicity provides a strong bottom up motivation for the Wino, in the sense that it represents a minimal extension to the Standard Model that solves the DM problem.
Accordingly, it is often motivated as a canonical minimal DM candidate~\cite{Cirelli:2005uq,Cirelli:2007xd,Cirelli:2008id,Cirelli:2009uv,Cirelli:2015bda}.
Yet, in addition to its simplicity, the Wino also emerge as a DM candidate in supersymmetric scenarios, see for example~\cite{Giudice:1998xp,Randall:1998uk,ArkaniHamed:2006mb}.

The Wino is also a testable DM candidate.
As it couples to the Standard Model through the weak force, the only free parameter is the mass.
If the Wino was produced as a thermal relic in the early universe, however, then even the mass becomes fixed at $2.9 \pm 0.1$ TeV~\cite{Hisano:2006nn,Hryczuk:2010zi,Beneke:2016ync}.
Then, as the Wino can annihilate to $\gamma \gamma$ and $\gamma Z$ at loop level, we would expect to be able to detect it using TeV gamma rays observed by Cherenkov telescopes, as we will demonstrate in the present work.

Nevertheless, the Wino indirect detection signature is significantly more involved than simply calculating the rate for $\chi^0 \chi^0 \to \gamma \gamma$ or $\gamma Z$ at one loop.
There are four additional effects which must be included in order to determine the observed photon flux to $\mathcal{O}(1)$.
The first of these is Sommerfeld enhancement, where the exchange of electroweak bosons between the triplet states of the Wino generates a potential, leading to a significant correction of the annihilation cross-section~\cite{Hisano:2003ec,Hisano:2004ds,Cirelli:2007xd,ArkaniHamed:2008qn}.
Secondly, given that $m_{\scriptscriptstyle {\rm DM}} \sim 3$ TeV is significantly larger than $m_{\scriptscriptstyle W}$, the cross talk between the DM mass and the electroweak scale generates large Sudakov double logarithms of the form $\alpha_{\scriptscriptstyle W} \ln^2 (m_{\scriptscriptstyle {\rm DM}}/m_{\scriptscriptstyle W})$, which must be resummed~\cite{Hryczuk:2011vi,Bauer:2014ula,Ovanesyan:2014fwa,Baumgart:2014saa,Baumgart:2015bpa,Ovanesyan:2016vkk}.
The third effect to include is the continuum emission of photons with energies well below $m_{\scriptscriptstyle {\rm DM}}$, arising from the decay of final state $W$ and $Z$ bosons.
These continuum photons can be included using public results for the spectra~\cite{Cirelli:2010xx}.
Finally, beyond line photons arising from two body decays, there can also be endpoint photons, which have energies near $m_{\scriptscriptstyle {\rm DM}}$, and can arise, for example, from three body decays~\cite{Baumgart:2015bpa,Baumgart:2017nsr,Baumgart:2018yed}.
Quantitatively, these photons have energies $E = z m_{\scriptscriptstyle {\rm DM}}$, and for $z \sim 1$, any instrument with finite resolution will be unable to distinguish these from the two-body contribution, and therefore they must be included.

In this work we will use the Wino photon spectrum and annihilation cross section that includes all of these effects, in particular we will use the results of~\cite{Baumgart:2018yed}.
Our results will be presented as a function of the cross section to line photons,
\begin{equation}
\langle \sigma v \rangle_{\rm line} = \langle \sigma v \rangle_{\gamma \gamma + \gamma Z/2}\,.
\end{equation}
In other words, this is the cross section associated with DM annihilating to two photons, plus half the cross section for $\gamma Z$, as the latter only provides a single photon.
The spectrum of photons resulting from this annihilation is then given by
\begin{equation}
\frac{dN_{\gamma}}{dE} = 2 \delta(E-m_{\scriptscriptstyle {\rm DM}}) + \frac{dN_{\gamma}^{\rm ep}}{dE} + \frac{dN_{\gamma}^{\rm ct}}{dE}\,.
\end{equation}
Here, the $\delta$ function corresponds to the two-body or line like photons, the analytic form of the endpoint (ep) spectrum can be found in~\cite{Baumgart:2018yed}, and for the continuum contribution we use the numerical results of PPPC 4 DM ID~\cite{Cirelli:2010xx}.

\begin{figure*}[tbh] 
\begin{center}
\includegraphics[width=0.48\textwidth]{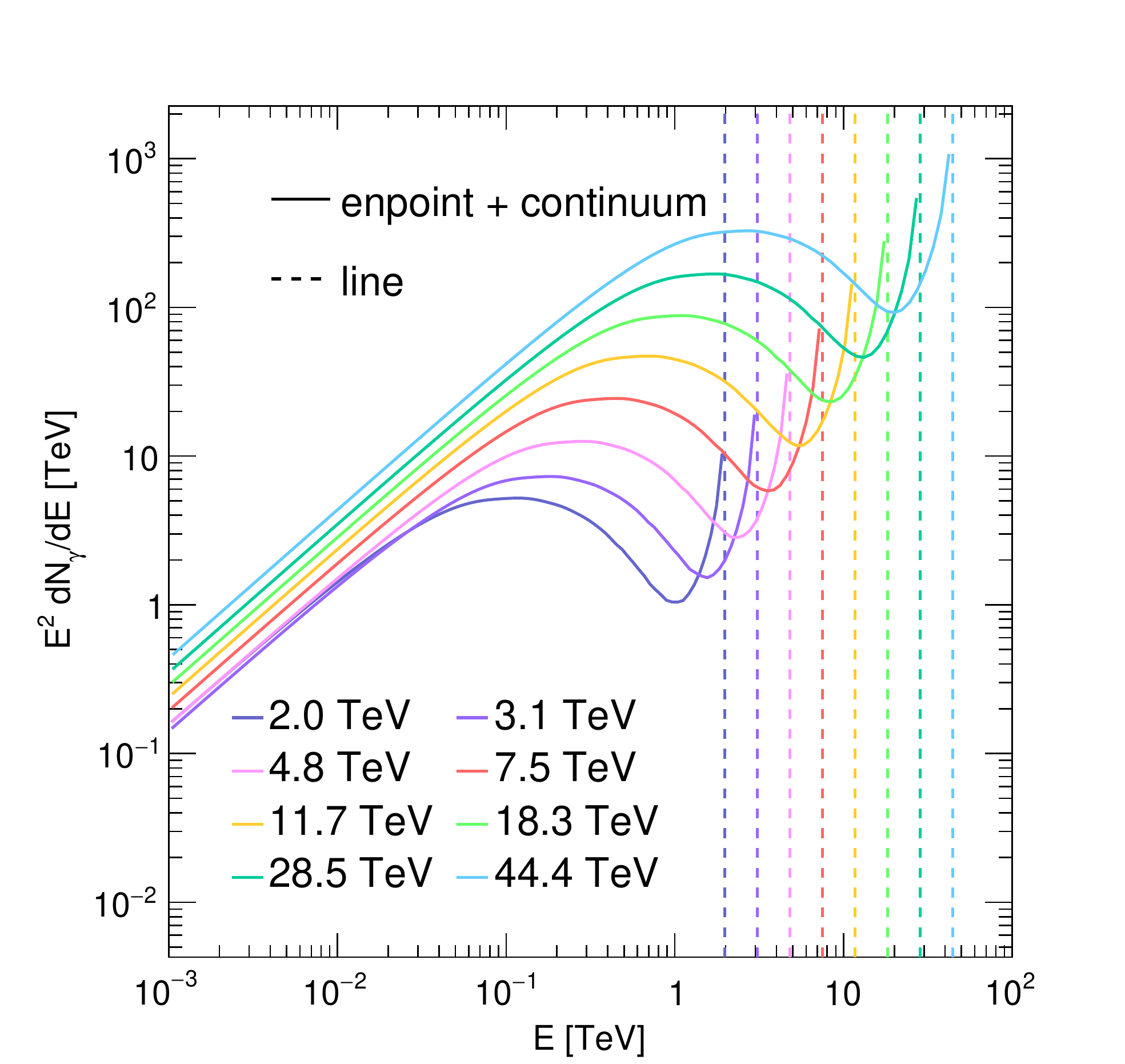}
\hfill
\includegraphics[width=0.48\textwidth]{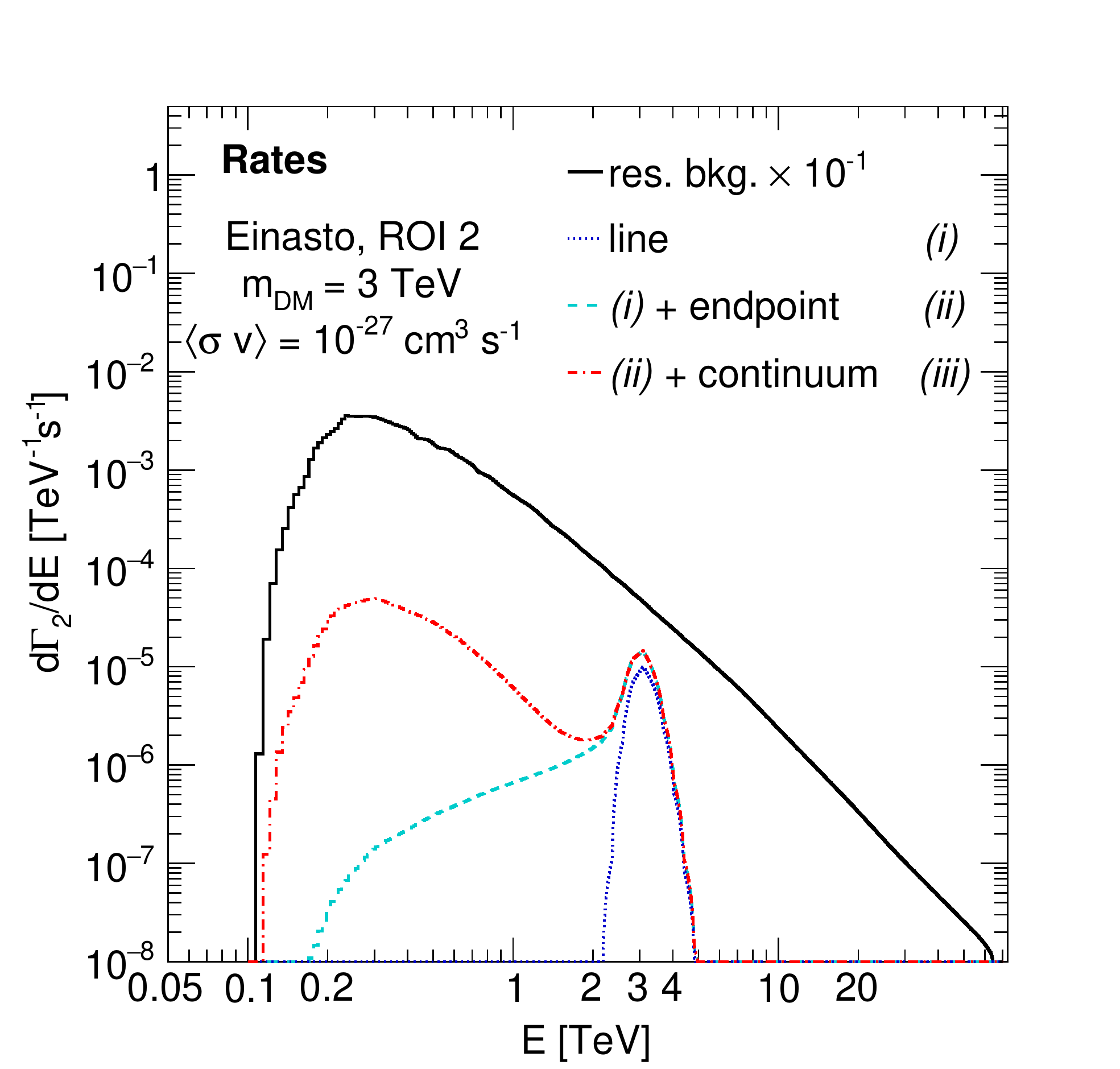}
\caption{{\it Left:} Wino spectrum for various DM masses. The endpoint and continuum contributions (solid lines) broaden the mono-energetic gamma-line spectrum (dashed lines). {\it Right:} Differential count rates in ROI 2 for H.E.S.S. mock observations of the GC. The residual background (black line) is shown as well as the Wino spectrum, assuming line-only contribution (blue line), line and endpoint (cyan line) and full spectrum including the continuum contribution (red line).}
\label{fig:fig1}
\end{center}
\end{figure*}

The left panel of Fig.~\ref{fig:fig1} shows for different DM masses the endpoint and continuum contribution to the Wino spectrum (solid lines) that are added to the gamma line (dashed lines).

\section{H.E.S.S. mock analysis framework}
\subsection{Dark matter density at the Galactic Center}

Using the theoretical results described above, we have in hand the cross-section for Wino DM to annihilate into photons, $\langle \sigma v \rangle_{\rm line}$, and further the spectrum that results from these annihilations, $dN_{\gamma}/dE$.
These quantities can then be used to determine the photon flux incident on the Earth from a region of the sky, which is given by
\begin{equation}
\frac{d\Phi_{\gamma}}{dE} = \frac{\langle \sigma v \rangle_{\rm line}}{8 \pi\, m_{\scriptscriptstyle {\rm DM}}^2} \frac{dN_{\gamma}}{dE} \times J\,.
\label{eq:flux}
\end{equation}
The additional quantity that must be specified in order to determine this flux is the $J$-factor, and is defined by
\begin{equation}
J = \int_{\rm ROI} d \Omega \int_0^{\infty} ds\,\rho_{\scriptscriptstyle {\rm DM}}^2\,,
\end{equation}
where $s$ is the line of sight between the observer and annihilation location. The region of interest (ROI) refers to the section of the sky being observed.
For our particular search we will consider observations of the center of the Milky Way, and the exact ROI will be defined in the next subsection.

Beyond the ROI, however, in order to obtain the $J$-factor we also need to specify a model for the distribution of DM within the Milky Way.
To this end, we adopt the Einasto profile~\cite{Einasto:1965czb}, specified as
\begin{equation}
\rho_{\rm Einasto}(r) = \rho_0 \exp \left[ - \frac{2}{\alpha} \left( \left( \frac{r}{r_s} \right)^{\alpha} - 1 \right) \right]\,.
\end{equation}
This profile is specified in terms of the Galactocentric radius $r$, and then we parameterize it taking $\alpha = 0.17$~\cite{Abramowski:2011hc}, $r_s = 20$ kpc~\cite{Pieri:2009je}, and finally the profile is normalized to agree with the local DM density.
In detail, we take $\rho_{\rm Einasto}(8.5~{\rm kpc}) = 0.39$ GeV/cm$^3$~\cite{Catena:2009mf}.

Of course, this single model for the DM distribution fails to encompass the present uncertainty that exists in its form, particularly near the GC.
For example, there is evidence to suggest that baryonic feedback could lead to the DM distribution within Milky-Way-like galaxies becoming cored at kpc scales~\cite{Chan:2015tna,Mollitor:2014ara}.
By this, we mean that the DM density $\rho_{\rm Einasto}(r)$ ceases to increase for Galactic radii below a core size $r_c$.
In this work we will show that the thermal Wino can be probed even for larger core sizes.

\subsection{H.E.S.S.-like observations and data analysis}
Mock observations of the GC region with H.E.S.S. are constructed based on the actual H.E.S.S.-I data taking in the region, {\it i.e.} including only the four small telescopes of the array. The region of interest for DM search is a disk of radius $1^\circ$ centered at the GC position. It is split in 7 sub-regions (ROIs) of width $0.1^\circ$. The main standard astrophysical emissions detected in VHE gamma rays in the field of view are excluded. A band up to latitudes $\pm0.3^\circ$ is excluded along the Galactic plane in order to reject the diffuse emission and the emissions from HESS J1745-290 and HESS J1747-281. A disk-shaped patch of radius $0.4^\circ$ is also excluded, centered at the position of HESS J1745-303.

The residual background in H.E.S.S. is made of hadrons that are misidentified as gamma rays and electrons that are difficult to distinguish from gamma rays.  The expected residual background in the ROIs is derived from the cosmic-ray proton, helium and electron spectra~\cite{Bernlohr:2012we}. In order to account for the characteristics of the detector and the observations, the spectra are convoluted for the H.E.S.S.-I Instrument Response Functions (IRFs), energy resolution (10\%) and effective area, and for the observation live time. The energy resolution is modeled as a Gaussian with with 0.1 and the H.E.S.S.-I effective area for observations at mean zenith angle of $20^\circ$ is used. A live time of 250 hours of observations is assumed to reproduce the data set used in Ref.~\cite{Abdallah:2018qtu}. In the case of cosmic-ray hadrons a rejection efficiency factor of 90\% is added. 
In order to compute the expected number of DM signal photons in the ROIs the flux in Eq.~\ref{eq:flux} is convoluted for the H.E.S.S.-I IRFs and the observations live time. The right panel of Fig.~\ref{fig:fig1} shows the differential count rate of the residual background and the Wino annihilation signal (line only, line and endpoint, full spectrum) in the ROI 2.

The computation of the sensitivity of H.E.S.S. to the expected DM annihilation signals is computed using a 2D-Poisson likelihood technique. The likelihood function contains a Poisson term for the signal region that contains the information about the expected signal, and a Poisson term for the control region where the residual background is determined. The likelihood function is binned in energy and space in order to exploit the peculiar spectral and spatial features of the DM signal that make it distinguishable from the power-law isotropic residual background. 
In order to test the hypothesis of presence of an expected signal from DM annihilation a log-likelihood ratio test statistics is performed. The test statistics is computed following Ref.~\cite{Cowan:2010js} for the case of limits calculation. A constraint at a 95\% confidence level (C.L.) on the free parameter of the likelihood, $\langle\sigma v\rangle_{\rm line}$, corresponds to a TS value of 2.71.

\section{Results}
\begin{figure*}[tbh] 
\begin{center}
\includegraphics[width=0.48\textwidth]{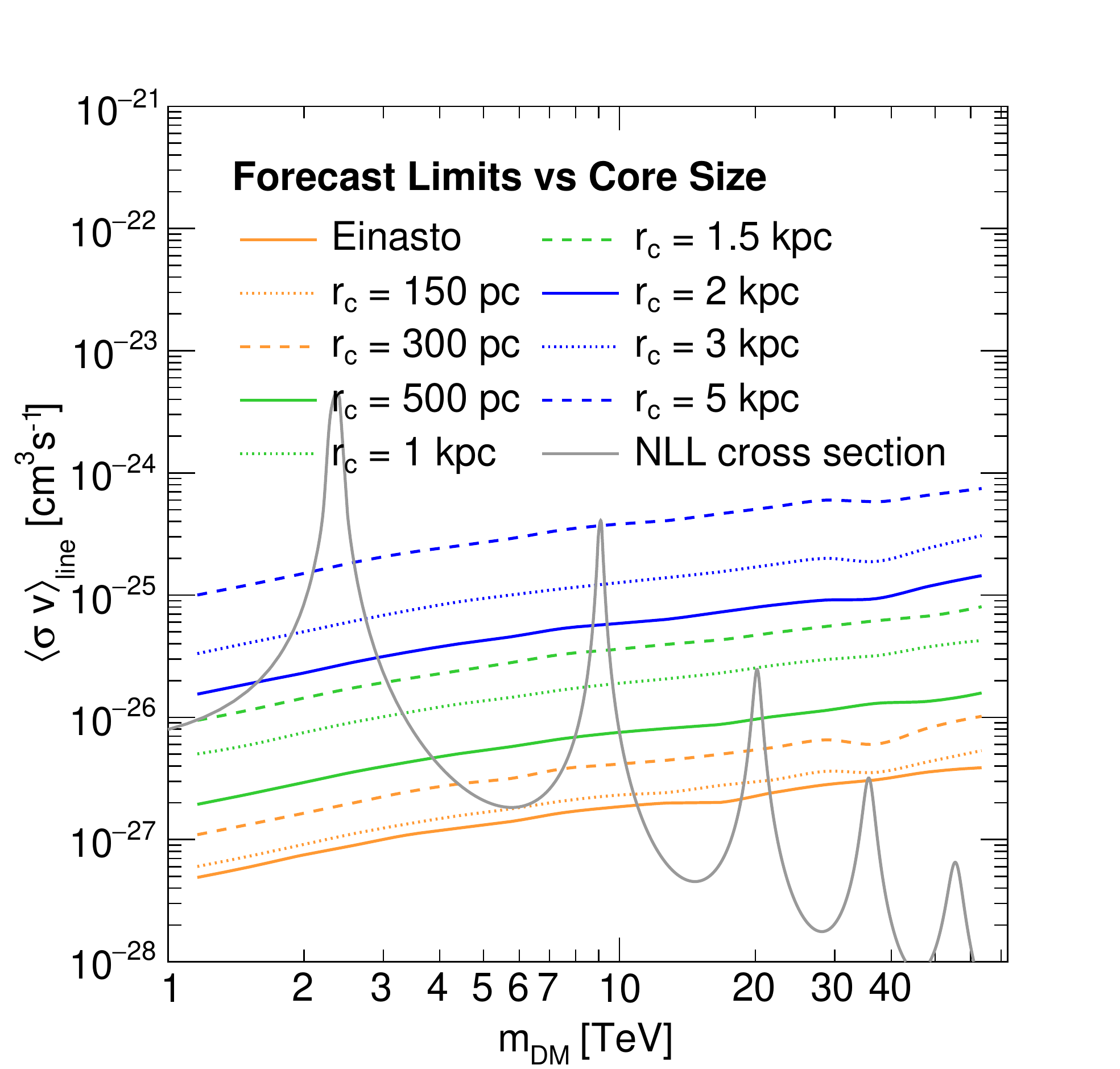}
\hfill
\includegraphics[width=0.48\textwidth]{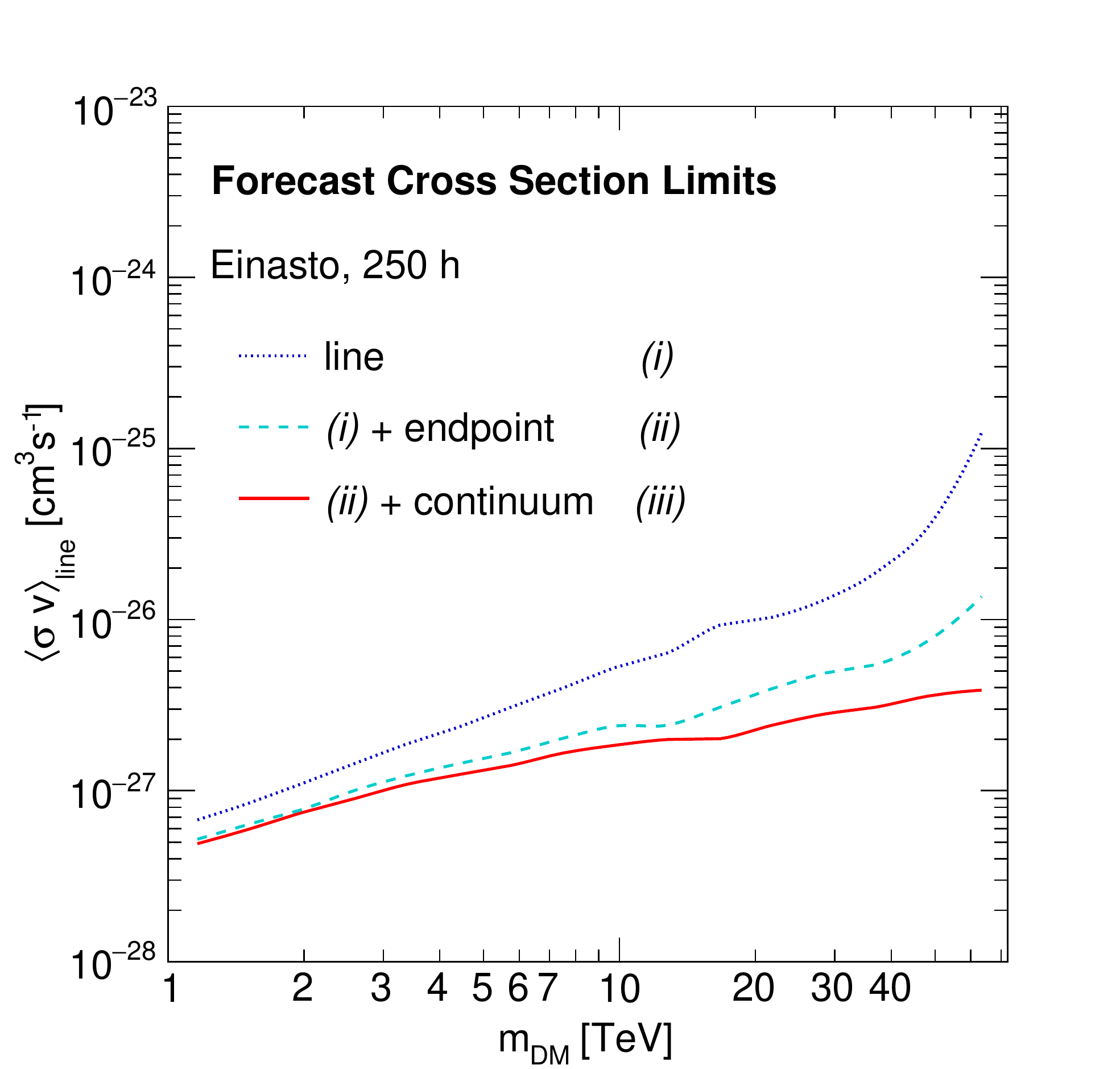}
\caption{H.E.S.S. 95\% C.L. upper limits on $\langle\sigma v\rangle_{\rm line}$ for Wino DM. {\it Left:} the limits are shown for the cuspy Einasto profile and for cored profiles of various sizes between 150~pc and 5~kpc. The predicted thermal relic cross section is shown (gray line). All the Wino masses for which the H.E.S.S. sensitivity is below the theoretical cross section are probed. {\it Right:} Comparison of the sensitivity to the line-only (blue line), line $+$ continuum (cyan line) and full (red line) Wino spectrum.}
\label{fig:fig2}
\end{center}
\end{figure*}

The sensitivity of H.E.S.S.~I to the Wino is shown in the left panel of Fig.~\ref{fig:fig2} and compared to the theoretical predicted cross section (gray) which represents the natural scale for DM annihilation and the values that the experiments aim to probe. Resonances in the predicted cross section are due to the Sommerfeld enhancement. The constraints are first computed assuming the Einasto profile. With this cuspy DM profile H.E.S.S.~I can probe the Wino DM candidate for masses up to about 10~TeV and at the position of the Sommerfeld resonances. This implies that H.E.S.S. is able to reject the Wino in most of the relevant DM mass range. However, there is large uncertainty on the DM profile at the GC. Thus, the sensitivity has been recomputed also for cored DM density profiles with cores of dimension: 150~pc, 300~pc, 500~pc, 1~kpc, 1.5~kpc, 2~kpc and 5~kpc. A significant degradation with respect to Einasto is observed, up to a factor about 200 for the largest core. Interestingly, Wino DM thermally produced in the early Universe with a mass $m_{\scriptscriptstyle{\rm DM}}=2.9$~TeV is probed for cores up to 2~kpc. The Wino is rejected for several-kpc cores (at the edge of what can be reasonably assumed) at the position of the resonance at 2.3~TeV.

The right panel of Fig.~\ref{fig:fig2} shows the impact of the components of the Wino spectrum, in addition to the mono-energetic gamma line due to the process $\chi^0\chi^0\to\gamma\gamma$. H.E.S.S.-I 95\% C.L. upper limits on $\langle\sigma v\rangle_{\rm line}$ are shown for the line only (dotted blue line), for the line $+$ endpoint contribution (dashed cyan line) and for the full spectrum including also the continuum contribution (solid red line). The major improvement comes from the endpoint contribution that has an impact on the limits of a factor 1.4 at 2.3, 1.5 at the thermal mass of 2.9~TeV and 2.1 at 9~TeV. The additional improvement due to the continuum contribution is of the 8\%, 12\% and 27\% at the same DM masses, respectively.

\section{Conclusions}
We built a framework for mock data analysis of H.E.S.S.-I observations of the GC region and computed the sensitivity to a Wino DM signal in the mass range between 1 and 70 TeV using the latest computations of the Wino spectrum. The H.E.S.S.-I telescope array rejects already the Wino as DM candidate in a mass range up to 10~TeV and at the position of the resonance at $\sim$20~TeV when assuming an Einasto profile. The H.E.S.S.-I sensitivity decreases significantly when assuming a cored DM profile, but the thermal Wino with $m_{\scriptscriptstyle\rm DM}=2.9$~TeV is rejected also for large cores. A further improvement is expected with H.E.S.S.~II using the full five-telescope hybrid array. Its sensitivity will benefit from the larger data set, the lower energy threshold due to the larger telescope CT5 and the new pointing strategy called Inner Galaxy Survey that extends the exposure of the observations to larger latitudes from the GC.

\section{Acknowledgement}
We thank Matt Baumgart, Timothy Cohen, Ian Moult,  Mikhail Solon, Iain Stewart, and Varun Vaidya for collaboration on the endpoint spectrum calculation. 
%
%
NLR is supported by the Miller Institute for Basic Research in Science at the University of California, Berkeley.
TRS is supported by the Office of High Energy Physics of the U.S. Department of Energy under grant Contract Numbers de-sc00012567 and de-sc0013999.

\end{document}